\documentclass[preprint,review,3p,12pt]{elsarticle}

\usepackage{amsmath}
\usepackage{graphics}
\usepackage{rotating}

\journal{Physica A}
\usepackage{amssymb}
\usepackage{setspace}
\usepackage[table]{xcolor}
\begin{document}


\begin{frontmatter}
\title{Quantifying individual performance in Cricket $-$ A network analysis of Batsmen and Bowlers   }
\author{Satyam Mukherjee }
\ead{satyam.mukherjee@gmail.com}
\address{Kellogg School of Management, Northwestern University, Evanston, Illinois 60208 USA}

\date{\today}     
\begin{abstract}
 Quantifying individual performance in the game of Cricket is critical for team selection in International matches. The number runs scored by batsmen and wickets taken by bowlers serves as a natural way of quantifying the performance of a cricketer.  Traditionally the batsmen and bowlers are rated on their batting or bowling average respectively. However in a game like Cricket it is always important the manner in which one scores the runs or claims a wicket. Scoring runs against a strong bowling line-up or delivering a brilliant performance against a team with strong batting line-up deserves more credit. A player's average is not able to capture this aspect of the game. In this paper we present a refined method to quantify the `quality' of runs scored by a batsman or wickets taken by a bowler. We explore the application of Social Network Analysis (SNA) to rate the players in a team performance. We generate directed and weighted network of batsmen-bowlers using the player-vs-player information available for Test cricket and ODI cricket. Additionally we generate network of batsmen and bowlers based on the dismissal record of batsmen in the history of cricket - Test ($1877-2011$) and ODI ($1971-2011$). Our results show that {\it M Muralitharan} is the most successful bowler in history of Cricket. Our approach could potentially be applied in domestic matches to judge a player's performance which in turn pave the way for a balanced team selection for International matches.                 

\end{abstract}



\end{frontmatter}

\section{Introduction}

Tools of Social Network Analysis (SNA) have been subject of  interest for theoretical as well as empirical study of social systems  \cite{laszlo,freeman,moviewatts}. A social network is a collection of people or groups interacting with each other and displaying complex features \cite{castellano}. Tools of SNA provide quantitative understanding for the human interaction of collective behavior. Considerable research has been done on scientific collaboration networks \cite{newman8, newman9,pan1,pan2}, board of directors, movie-actor collaboration network \cite{moviewatts} and citation networks \cite{solla,redner,bergstrom,bergstrom2,vespignani}. The use of network analysis not only provides a global view of the system, it also shows the complete list of interactions. In the world of sports, individual players interact with each other and also with the players in the opponent team. It is therefore important to study the effect of interactions on performance of a player.

In recent years there has been an increase in study of quantitative analysis of individual performance involving team sports. Time series analysis have been applied to football \cite{naim05,bittner}, baseball \cite{petersen08,sire09}, basketball \cite{naim07,skinner10,guerra} and soccer \cite{Rubner,Ribeiro}. Quantifying the individual performance or `quality' of a player in any sport is a matter of great importance for the selection of team members in international competitions and is a topic of recent interest \cite{bhandari,condon}. A lot of negotiations are involved in the process of team-selection \cite{sharda}. Studies have focussed on non-linear modeling techniques like neural networks to rate an individual's performance.  For example, neural networks techniques were used to predict the performance of individual cricketer's based on their past performance \cite{sharda}. Earlier tools of neural networks were used to model performance and rank NCAA college football teams \cite{wilson1995},  predicting javelin flights \cite{majer}and to recognize patterns in Table Tennis and Rowing \cite{jurgen}. 
Again, a model-free approach was developed to extract the outcome of a soccer match\cite{heuer10}. It was also shown that the statistics of ball touches presents power-law tails and can be described by $q$-gamma distributions \cite{malacarne}. In recent years, the study of complex networks have attracted a lot of research interests \cite{laszlo}. The tools of complex network analysis have previously been applied to quantify individual brilliance in sports and also to rank the individuals based on their performance. For example, a network approach was developed to quantify the performance of individual players in soccer \cite{duch10}. Network analysis tools have been applied to football\cite{yokoyama} and Brazilian soccer players \cite{onody04}. Successful and un-successful performance in water polo have been quantified using a network-based approach \cite{mendes2011}. Head-to-head matchups between Major League Baseball pitchers and batters was studied as a bipartite network \cite{saavedra09}. More recently a network-based approach was developed to rank US college football teams \cite{newman2005}, tennis players \cite{radicchi11} and cricket teams and captains \cite{mukherjee2012}. 

The complex features of numerous social systems are embedded in the inherent connectivity among system components \cite{laszlo,mendes2011}. Social network analysis (SNA) provides insight about the pattern of interaction among players and how it affects the success of a team \cite{Lusher}. This article points out that how topological relations between players help better understanding of individuals who play for their teams and thus elucidate the individual importance and impact of a player. In this paper we apply the tools of network analysis to batsmen and bowlers in cricket and quantify the `quality' of an individual player. The advantage of network based approach is that it provides a different perspective for judging the excellence of a player. 

We take the case of individual performance of batsmen and bowlers in International Cricket matches.  Cricket is a game played in most of the Commonwealth countries. The International Cricket Council (ICC) is the government body which controls the cricketing events around the globe. Although ICC includes $120$ member countries, only ten countries with `Test' status - Australia, England, India, South Africa, New Zealand, West Indies, Bangladesh, Zimbabwe, Pakistan and Sri Lanka play the game extensively. There are three versions of the game - `Test', One Day International (ODI) and Twenty20 (T20) formats. Test cricket is the longest format of the game dating back to $1877$. Usually it lasts for five days involving $30-35$ hours. Shorter formats, lasting almost $8$ hours like ODI started in $1971$ and during late $2000$ ICC introduced the shortest format called T20 cricket which lasts approximately $3$ hours \cite{amy2007}.  

Batsmen and Bowlers in Cricket are traditionally ranked according to their batting and bowling average respectively. Judged by the batting average, Sir Donald Bradman (with an average of $99.94$) is regarded as the greatest batsman of all times. The next best batting average of $60.9$ is held by Graeme Pollock. Even though most of the records held by Bradman has been eclipsed by modern day batsmen like Sachin Tendulkar, Brian Lara, Graham Gooch, Mohammad Yusuf, Bradman's legacy still survives and generates debate among fans about his greatness relative to recent players like Sir Vivian Richards, Brian Lara or Sachin Tendulkar. The question thus naturally arises is whether batting average of batsmen (or bowling average of bowlers) are the best measure for judging the worth of a batsman (or a bowler). It was shown that rankings based on average suffer from two defects - $i)$ Consistency of scores across innings and $ii)$ Value of runs scored by the player \cite{Vani2010}. However one should also consider the quality of bowling as well. For example according to Bradman himself, the greatest innings he ever witnessed was that of McCabe's innings of $187$ at Sydney in 1932. The reason being it came against Douglas Jardine's body-line attack, widely regarded as one of the fiercest bowling attacks.  Similarly runs scored against West Indian bowlers like Michael Holding, Joel Garner, Malcom Marshall and Andy Roberts deserve more credit than runs scored against low bowling attack of Bangaldesh or Zimbabwe. On similar arguments the wicket of top-order batsman is valued more than the wicket of a lower-order batsman.   If a bowler claims the wicket of Bradman, Lara, Richards or Tendulkar, he gets more credit than if he dismiss any lower-order batsman. Under the usual ranking scheme based on bowling average, {\it George Lohmann } of England has the lowest (best) bowling average ($10.75$) in Test cricket. However bowlers like {\it George Lohmann} played under pitch conditions favoring fast bowlers. Hence batting (or bowling) average does not serve as an efficient gauge for a batsman's (or bowler's) ability \cite{lemmer}. Against, this background, we propose a network based approach to quantify the `quality' of a batsman or bowler. The rest of the paper is presented as follows : In Section 2 we propose the methods of link formation among the batsmen and bowlers. In section 3 we discuss the results and we conclude in Section 4. 

\section{Methodology}

We obtain data from the cricinfo website \cite{cricinfo}. The website contains the information of proceedings of all Test matches played since $1877$ and all ODI matches from $1971$ onwards. These include the runs scored by batsmen, wickets taken by bowlers, outcome of a game and also the information of the mode of dismissal of a batsman. We collect the data of player-vs-player for Test cricket ($2001-2011$), ODI cricket ($1999-2011$)  from the cricinfo website. The data of player-vs-player contains the information of runs scored by a batsman against every bowler he faced and how many times he was dismissed by the bowlers he faced. No information of player-vs-player is available for games played earlier than $2001$. We also collect the batting and bowling averages of players from the player's profile available in the cricinfo website. Batting average of a batsman is defined as the total number of runs scored by the batsman divided by the number of times he was dismissed. Thus higher batting average reflects higher `quality' of a batsman. Similarly, bowling average is defined as the number of runs given by the bowler divided by the number of wickets claimed by him. Thus lower bowling average indicates higher ability of the bowler. This information is used to generate the network of interaction among bowlers and batsmen in cricket matches. 

\subsection{Weighted and Directed Network}

Cricket is a bat-and-ball game played between two teams of $11$ players each. The team batting first tries to score as many runs as possible, while the other team bowls and fields, trying to dismiss the batsmen. At the end of an innings, the teams switch between batting and fielding. This can be represented as a directed network of interaction of batsmen ($B_{a}$) and bowlers ($B_{o}$). Every node in $B_{o}$ has a directed link to all nodes in $B_{a}$, provided the batsman and bowler face each other. The performance of a batsman is judged by the `quality' of runs scored and not the number of runs scored. Hence runs scored against a bowler with lower bowling average carries more credit than runs scored against a bowler of less importance. We introduce a performance index of a batsman ($PIB$) against a bowler given by the following equation

\begin{equation}
PIB = \frac{A_{Ba}}{C_{Bo}}
\end{equation}
 where $A_{Ba}$ is the batting average of the batsman against the bowler he faced and $C_{Bo}$ refers to the career bowling average of the bowler. Mathematically, batting average of the batsman ($A_{Ba}$) is given by the ratio $\frac{R}{d}$ where $R$ is the number of runs scored against a bowler and $d$ is the number of times he was dismissed by the bowler \footnote{$R$ and $d$ are evaluated for Test matches played between $2001$ and $2011$ and ODI ($1999-2011$)}. Hence, if  the career bowling average of a bowler is low (indicating a good bowler), $PIB$ increases indicating that the batsman scored runs against quality opposition. We generate a weighted and directed network of bowlers to batsmen where weight of the link is given by  $PIB$. The network generated is thus based on the directed interaction of $B_{o}$  and $B_{a}$. For the weighted network the in-strength $s_{i}^{in}$ is defined as 
 
\begin{equation}
s_{i}^{in} = \displaystyle\sum_{j \ne i} W_{ji}
\end{equation}
where $W_{ji}$ is given by the weight of the directed link.

So far, we have concentrated on the performance index of batsmen since $2001$. Although the data for player-vs-player is not available for dates earlier than $2001$,  one could quantify the overall performance of a bowler based on the dismissal record of batsmen. For example, the wicket of a top-order batsman always deserve more credit than the wicket of a tail-ender. Thus the 'quality' of dismissal serves as a measure for the greatness of a bowler. We define the quality index of bowler ($QIB$) as 

\begin{equation}
QIB = D \frac{C_{Ba}}{C_{Bo}}
\end{equation}
where $D$ is defined as the number of times a batsman was dismissed by a particular bowler, $C_{Ba}$ refers to the career batting average of a batsman and $C_{Bo}$ indicates the career bowling average of a bowler.  Thus a greater value of $QIB$ indicates a better rank of a bowler. As before, we construct weighted and directed networks, this time the directed link pointing towards the bowlers. We evaluate the in-strength of the bowlers, which serves as a quantification of the `quality' of a bowler. 

\subsection{Projected network}

\begin{figure*}
\begin{center}
\includegraphics[width=9.0cm]{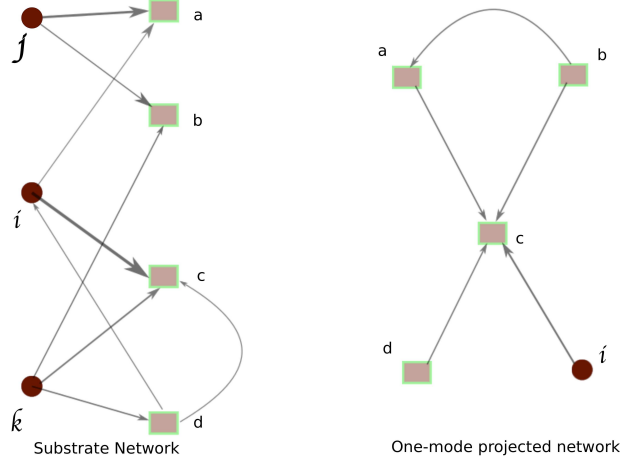}
\caption{ \label{fig:network00}(Color online) Substrate network of batsmen and bowlers. The thickness of the directed link is proportional to the $QIB$. The resultant network of bowlers is constructed if the bowlers dismiss the same batsman and they are contemporary players. The direction and weights of the links are applied according to the gradient scheme of link formation. }
\end{center}
\end{figure*}

\begin{figure*}
\begin{center}
\includegraphics[width=9.0cm]{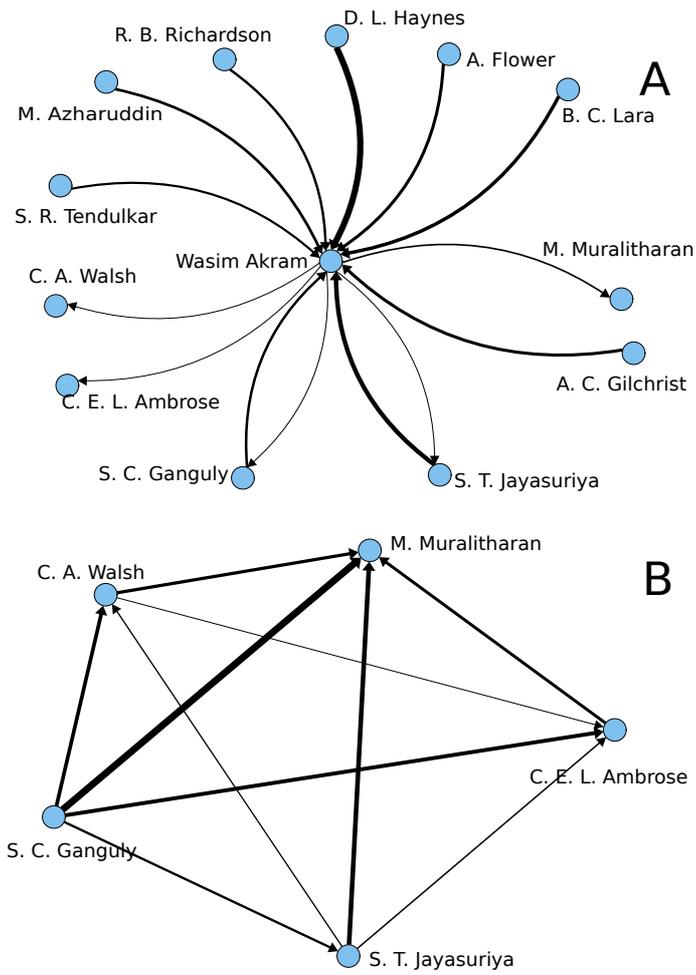}
\caption{ \label{fig:network0}(Color online) (A) Subgraph of the substrate network of batsmen and bowlers in ODI ($1971-2011$). The thickness of the directed link is proportional to the $QIB$. (B) The resultant projected network of bowlers is constructed if the bowlers dismiss the same batsman (here it is {\it Wasim Akram}).}
\end{center}
\end{figure*}

The manner in which the game is played doesn't allow us to compare the relative dominance of one batsman over another batsman or one bowler over another bowler. Unlike in tennis, where each player has to compete directly with the opponent, in cricket a batsman is pitted against a bowler. Hence it is very difficult to judge the relative superiority of a batsman (bowler) over another batsman (bowler). The in-strength of a bowler or batsman conveys the `quality' of dismissal by a bowler or the `quality' of runs scored by a batsman. However, it doesn't reflect the relative importance or popularity of one player over other players.  To address this issue, in this section we generate one-mode projected network between batsmen who face the same bowler (or bowlers who dismiss the same batsman) in which the links are generated according to the method of gradient link formation.  Traditionally a gradient network is constructed as follows. Consider a substrate network $S$. Each node $i$ in the network is assigned with a random number $h_{i}$ which describes the `potential' of the node. Gradient network is constructed by directed links that point from each node to the nearest neighbor with highest potential \cite{danila, jamming}. Here we take a slightly different route to construct the projected network.

 In Figure~\ref{fig:network00} we demonstrate the generation of the one-mode projected network according to the gradient scheme of link formation. First we consider the substrate network of batsmen and bowlers according to the dismissal records. The thickness of the edge is proportional to $QIB$. Thus if batsman $i$ is dismissed by bowlers {\tt a} and {\tt c}, then bowlers {\tt a} and {\tt c} are connected. We evaluate the in-strength $s_{i}^{in}$ of the nodes {\tt a} and {\tt c}. The in-strength acts a `potential' for each bowler. We construct gradient links between two bowlers along the steepest ascent, where the weight of the directed link is the difference of the in-strength of two nodes. Thus weighted and directed links are formed between two bowlers if they dismiss the same batsman. We repeat this procedure for all the nodes in the substrate network and a resultant one-mode projected network is formed. Additionally we introduce a constraint, in which two bowlers are linked only if they are contemporary. Thus {\tt b} and {\tt d} are not linked in the gradient scheme since they are not contemporary players. We apply the same method of gradient link formation on batsmen, where the weight of each link in the substrate network is proportional to the $PIB$. The weight ${\omega}_{ij}$ of a gradient-link is given as

\begin{equation}
\omega_{ij} = |s_{i}^{in} - s_{j}^{in}| 
\end{equation}

where $s_{i,j}^{in}$ are the in-strength of two nodes $i$ and $j$. The projected network thus highlights the relative importance of a player over other. We construct the substrate network of batsmen and bowlers for Test cricket and ODI cricket and construct the projected network of players. Next we apply the PageRank algorithm on the resultant projected network and evaluate the importance of each player. In Figure~\ref{fig:network0}(A) we show a subgraph of the substrate network of batsmen and bowlers in ODI ($1971-2011$). The projected network of bowlers is generated if they dismiss the same batsman ({\it Wasim Akram}) (See Figure~\ref{fig:network0}(B) ). In the same way one can construct projected network of batsmen who are dismissed by {\it Wasim Akram}.

\subsubsection{PageRank algorithm}

We quantify the importance or `popularity' of a player with the use of a complex network approach and evaluating the PageRank score, originally developed by Brin and Page \cite{brinpage}. Mathematically, the process is described by the system of coupled equations

\begin{equation}
    p_i =  \left(1-q\right) \sum_j \, p_j \, \frac{{\omega}_{ij}}{s_j^{\textrm{out}}}
+ \frac{q}{N} + \frac{1-q}{N} \sum_j \, \delta \left(s_j^{\textrm{out}}\right) \;\; ,
\label{eq:pg}
\end{equation}
where ${\omega}_{ij}$ is the weight of a link and $s_{j}^{out}$ = $\Sigma_{i} {\omega}_{ij}$ is the out-strength of a link. $p_i$ is the PageRank score assigned to team $i$ and represents the fraction of the overall ``influence'' sitting in the steady state of the diffusion process on vertex $i$ (\cite{radicchi11}).  $q \in \left[0,1\right]$ is a control parameter that  awards a `free' popularity to each player and $N$ is the total number of players in the network. 
The term $ \left(1-q\right) \, \sum_j \, p_j \, \frac{{\omega}_{ij}}{s_j^{\textrm{out}}}$  represents the portion of the score received by node $i$ in the diffusion process obeying the hypothesis that nodes  redistribute their entire credit  to neighboring nodes. The term $\frac{q}{N}$ stands for a uniform redistribution of credit among all nodes. The term $\frac{1-q}{N} \, \sum_j \, p_j \, \delta\left(s_j^{\textrm{out}}\right)$ serves as a correction in the case of the existence nodes with null out-degree, which otherwise would behave as sinks in the diffusion process.  It is to be noted that the PageRank score of a player depends on the scores of all other players and needs to be evaluated at the same time. To implement the PageRank algorithm in the directed and weighted network, we start with a uniform probability density equal to $\frac{1}{N}$ at each node of the network. Next we iterate through  Eq.~(\ref{eq:pg}) and obtain a steady-state set of PageRank scores for each node of the network. Finally, the values of the PageRank score are sorted to determine the rank of each player. According to tradition, we use a uniform value of $q=0.15$. This choice of $q$ ensures a higher value of PageRank scores \cite{radicchi11}.  In general it is difficult to get analytical solutions for Eq.~(\ref{eq:pg}) \cite{radicchi11, easley}. Although in the simplest case of a single tournament an analytical solution for values of $p_{i}$ was determined \cite{radicchi11}, in Cricket such a situation is not possible since it is a team game. The values of $p_i$s are evaluated recursively by setting $p_{i}=\frac{1}{N}$. Then we iterate Eq.~(\ref{eq:pg}) until a steady-state of values is reached.

\section{Results}

\begin{figure*}
\begin{center}
\includegraphics[height=6cm]{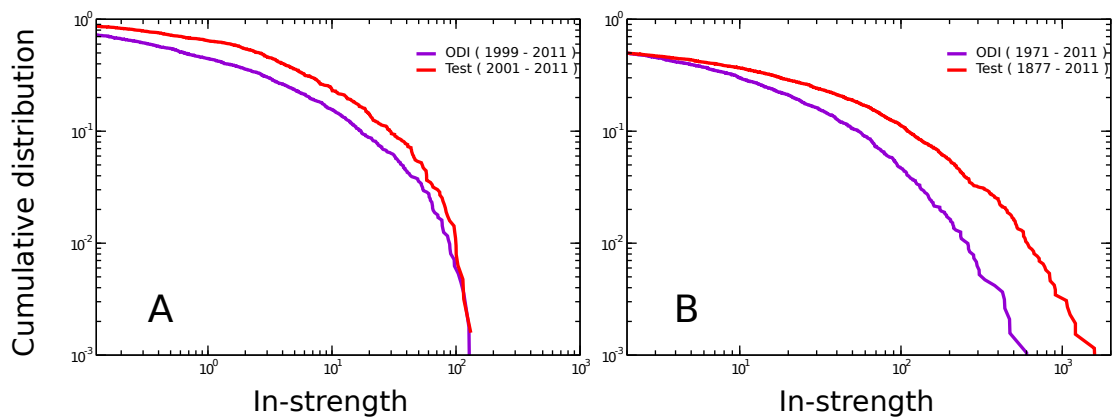}
\caption{ \label{fig:network2}(Color online) In-strength distribution of the weighted and directed network of (A) batsmen in Test cricket ($2001 - 2011$) and ODI cricket ($1999 - 2011$) and (B) bowlers network in the history of Test cricket ($1877-2011$) and ODI ($1971-2011$). }
\end{center}d
\end{figure*}

In this section, we explore the in-strength distribution of the weighted and directed networks. The in-strength of a node is an indication of the performance of an individual against the opponent team member. Thus a greater value of in-strength indicates a better the performance of the individual. In Fig~\ref{fig:network2} we plot the cumulative in-strength distribution of batsmen and bowlers in Test cricket  and ODI cricket. The in-strength distribution reflects the topology of the network and how the players interact with each other. As show in Fig~\ref{fig:network2}(A), the in-strength distribution decays slowly for smaller values of in-strength ($\approx 70$). For values higher than $70$, the in-strength distribution decays at a much faster rate. This is in contrast with the in-strength distribution of bowlers (Fig~\ref{fig:network2}(B)), where the decay is slow. The reason being that all bowlers have to bat once the top order batsmen have been dismissed, thus establishing more links for the batsmen. However not all batsmen are specialist bowlers, which leads to low connections for bowlers.  

\begin{figure*}
\begin{center}
\includegraphics[height=10cm]{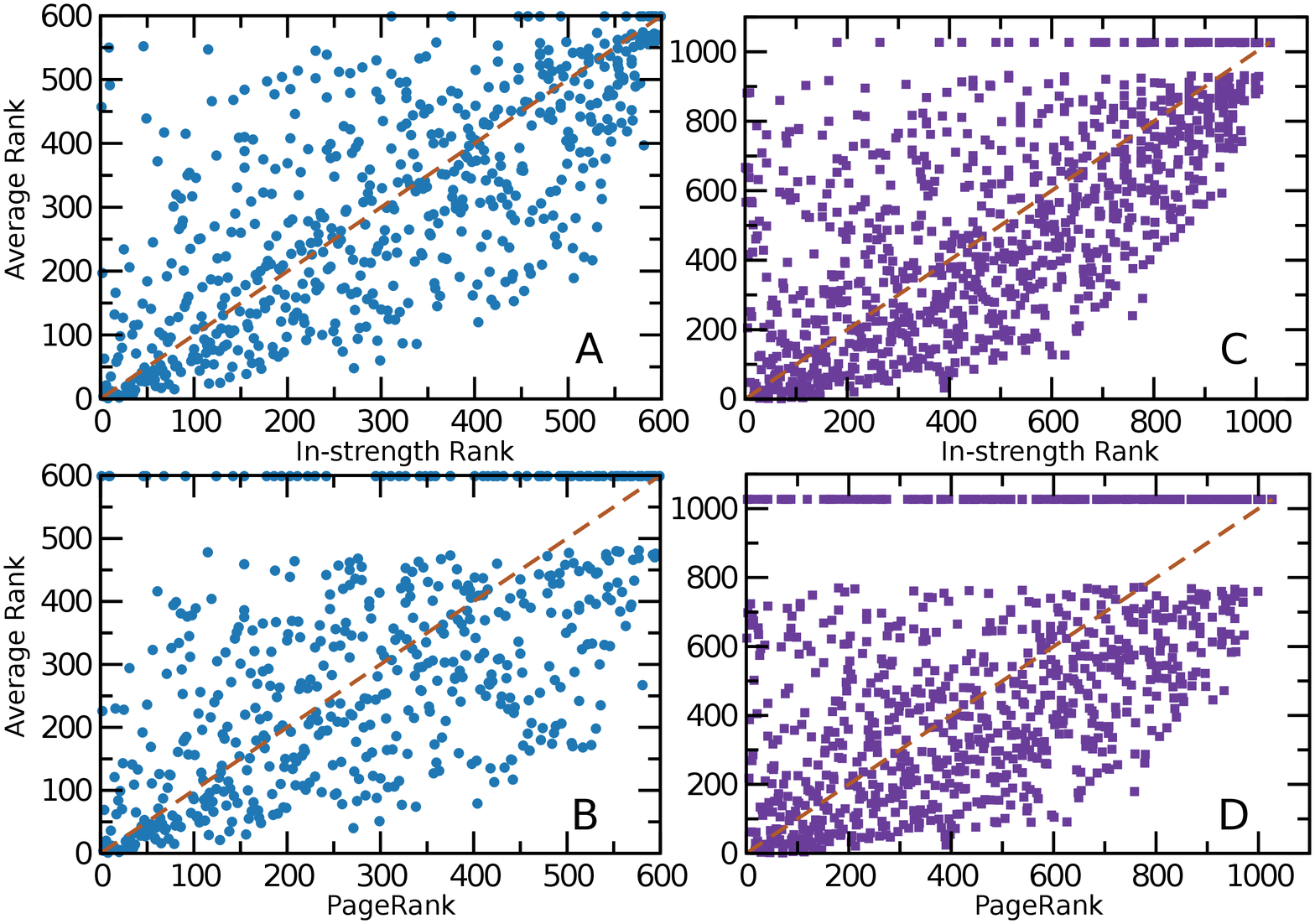}
\caption{ \label{fig:corr1}(Color online) (A) Scatter plot of between the rank positions obtained according to batting average rank and In-strength rank for Test cricket ($2001-2011$) ;  Spearman correlation $\rho$ = $0.71$.  (B) Scatter plot of between the rank positions obtained according to batting average rank and PageRank score for Test cricket ($2001-2011$) ; Spearman correlation $\rho$ = $0.62$. (C) Scatter plot of between the rank positions obtained according to batting average rank and In-strength rank for ODI cricket ($1999-2011$) ; Spearman correlation $\rho$ = $0.69$. (D) Scatter plot of between the rank positions obtained according to batting average rank and PageRank score in ODI cricket ($1999-2011$) ; Spearman correlation $\rho$ = $0.61$.}
\end{center}
\end{figure*}

As mentioned above the in-strength of a batsman reflects the performance of a batsman in terms of quality of runs scored. In Table~\ref{tab:table1} we list the top $50$ batsmen in Test cricket between $2001$ and $2011$. The batsmen are ranked according to their in-strength.  We observe that {\it K. C. Sangakkara} of Sri Lanka occupies the top spot followed by India's {\it S. R. Tendulkar} with Australia's {\it R. T. Ponting}  and  South Africa's {\it J. H. Kallis} occupying the third and fourth spot respectively. {\it R. Dravid} of India occupies the fifth position. We compare the in-strength rank with the PageRank score and batting average of batsmen for runs scored between $2001$ and $2011$. Additionally we list the best ever cricket rating received by a batsman between $2001$ and $2011$. {\bf In Figure~\ref{fig:corr1}(A, B) we compare the correlation of ranks obtained from in-strength and PageRank algorithm with batting average. We observe that ranks obtained from batting average is positively correlated with in-strength rank and PageRank score}.
Judged by the batting average and the ICC points we observe that {\it B. C. Lara} of West Indies emerge as the most successful batsman in Test cricket between $2001$ and $2011$. Similarly Australia's {\it R. T. Ponting} averages more than {\it S. R. Tendulkar} and {\it K. C. Sangakkara}. However both {\it K. C. Sangakkara } and {\it S. R. Tendulkar} accumulated runs against better bowling attack. 
In Table~\ref{tab:table2} we list the top $50$ batsmen in ODI cricket ($1999-2011$). {\bf As shown in Figure~\ref{fig:corr1}(C, D) we observe that ranks obtained from batting average is positively correlated with in-strength rank and PageRank score. The top $5$ positions according to in-strength rank or PageRank do not correspond with that of batting average or ICC rankings.}   Again {\it K. C. Sangakkara} emerge as the most successful batsman followed by Australia's {\it R. T. Ponting } and India's {\it S. R. Tendulkar}. Even though {\it S. R. Tendulkar} averages more than his predecessors and also received the highest ICC points, both {\it K. C. Sangakkara } and {\it R. T. Ponting } scored runs against better bowling attack. . Note that this ranking is sensitive to change in information of player-vs-player once the information prior to the year $2000$ is available in the cricinfo website. 

\begin{figure*}
\begin{center}
\includegraphics[height=10cm]{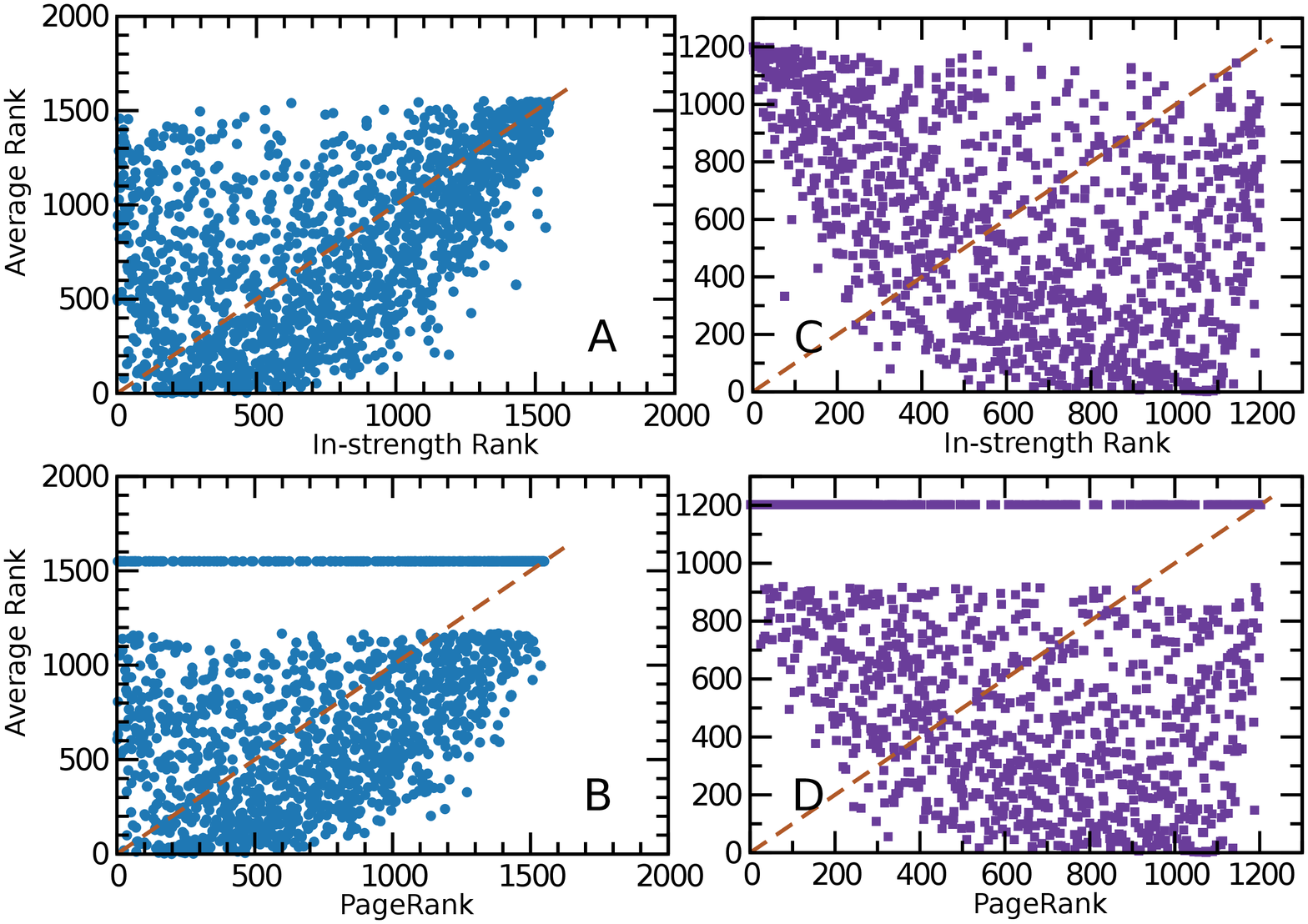}
\caption{ \label{fig:corr2}(Color online) (A) Scatter plot of between the rank positions obtained according to bowling average rank and In-strength rank for Test cricket ($1877-2011$) ; Spearman correlation $\rho$ = $0.53$.  (B) Scatter plot of between the rank positions obtained according to bowling average rank and PageRank score for Test cricket ($1877-2011$) ; Spearman correlation $\rho$ = $0.46$. (C) Scatter plot of between the rank positions obtained according to bowing average rank and In-strength rank for ODI cricket ($1971-2011$) ; Spearman correlation $\rho$ = $-0.44$. (D) Scatter plot of between the rank positions obtained according to bowling average rank and PageRank score in ODI cricket ($1971-2011$) ; Spearman correlation $\rho$ = $-0.34$.}
\end{center}
\end{figure*}

{\bf We rank the performance of all bowlers in Test cricket ($1877-2011$) in Table~\ref{tab:table3}, and identify bowlers with highest influence. We observe that the bowlers ranked by the average are different from that obtained from SNA. In  Figure~\ref{fig:corr2}(A, B) we compare the ranking obtained from in-strength and PageRank algorithm with bowling average. We observe a low positive correlation between the different ranking schemes.} We observe that according to in-strength values Sri Lanka's {\it M. Muralitharan} emerge as the most successful bowler in the history of Test cricket ($1877-2011$) followed by {\it S. K. Warne} (AUS), {\it G. D. McGrath} (AUS), {\it A. Kumble} (IND) and {\it C. A. Walsh} (WI) (See Table~\ref{tab:table3}). As before we generate gradient network of bowlers and apply the PageRank algorithm. It is interesting to note that the top five bowlers according to PageRank score are {\it M. Muralitharan} (SL), {\it S. K. Warne} (AUS), {\it G. D. McGrath} (AUS), {\it F. S. Trueman} (ENG) and {\it C. A. Walsh} (WI) (See Table~\ref{tab:table3}). Thus according to quality of `dismissal' and relative `popularity' of bowlers {\it M. Muralitharan} emerge as the most successful bowler in Test cricket. Interestingly, {\it M. Muralitharan} is the highest wicket-taker in Test cricket. His success could be {\it a posteriori} justified by his long and successful career spanning $18$ years (between $1992$ and $2010$). During his entire career {\it M. Muralitharan} dismissed $800$ batsmen (highest in Test cricket) which included the likes of {\it S. R. Tendulkar} (dismissed $14$ times), {\it R. Dravid} (dismissed $12$ times) and {\it B. C. Lara} (dismissed $9$ times). In addition to this he holds the record of maximum number of five wickets in an innings ($67$ times) and ten wickets in a match ($22$ times). We also observe that {\it S. K. Warne}, the second best bowler in Test cricket has second highest number of dismissals ($708$) to his credit. Both these bowlers had extremely long and successful careers spanning almost two decades. Australia's {\it G. D. McGrath}, who has been considered one of the best fast bowlers in cricket holds a better average than that of his immediate predecessors. However his in-strength rank and PageRank score indicates that his quality of dismissal were not better than {\it Muralitharan } or {\it Warne}. This leads to the possible question - are bowling averages the best indicator of a bowler's ability ?. In our all time top $50$ list we observe that England's {\it S. F. Barnes} has the best bowling average of $16.43$ and highest ICC points of $932$ among all the bowlers (as listed in Table~\ref{tab:table3}). However like {\it George Lohmann}, {\it S. F. Barnes} too enjoyed favorable pitch conditions. The batsmen playing in such pitches usually averaged lower than the recent batsmen. Hence for players like {\it S. F. Barnes}, the $QIB$ is low which in turn affects his in-strength. However, his PageRank score his higher than most of the modern age bowlers indicating his relative `popularity' or supremacy over other bowlers. A similar situation is seen with Pakistan's {\it Imran Khan}. Although his in-strength is lower than that of {\it Wasim Akram} or {\it D. K. Lillee}, his PageRank score is higher than most of his predecessors. Rankings based on SNA show little agreement with traditional methods of performance evaluation. 
 
In ODI history ($1971-2011$) too, Sri Lanka's {\it M. Muralitharan} leads the list of top $50$ bowlers, followed by Pakistan's {\it Wasim Akram}, Australia's {\it G. D. McGrath}, Pakistan's {\it Waqar Younis} and South Africa's {\it S. M. Pollock}. PageRank scores reveal that {\it M. Muralitharan} is the most successful bowler followed by {\it Wasim Akram} (PAK), {\it Waqar Younis} (PAK), {\it G. D. McGrath} (AUS) and{\it B. Lee} (AUS).  Although {\it G. D. McGrath} has a slightly better average than {\it M. Muralitharan}, he falls short of the latter in terms of in-strength, PageRank score and ICC points. Again, judged by the number of dismissals, {\it M. Muralitharan} heads the list with $534$ wickets, with {\it Wasim Akram} and {\it Waqar Younis } occupying the second and third position respectively. There are few surprises in the list. India's {\it A. B. Agarkar} is placed above in comparison to {\it N. Kapil Dev} (IND), {\it C. E. L. Ambrose} (WI) or {\it C. A. Walsh} (WI) whom cricket experts consider as better bowlers. However, what goes in favor of {\it A. B. Agarkar} is the `quality' of wickets he took. Thus even though he went for runs and didn't have a long career, he was able to dismiss most of the batsmen with good average. {\bf In Figure~\ref{fig:corr2}(C, D) we compare the ranks obtained from in-strength and PageRank with bowling average. We observe that ranking schemes obtained from PageRank (and in-strength) are anti-correlated with the bowling average.} This is not surprising in the sense that bowling average is not a proper way of judging a player's performance. Also in the ODIs, there has been a practice of bringing in part-time bowlers who have low-averages. This is paradoxical in the sense that it indicates part-time bowlers are better than the regular bowlers. 

\begin{figure*}
\begin{center}
\includegraphics[height=10cm]{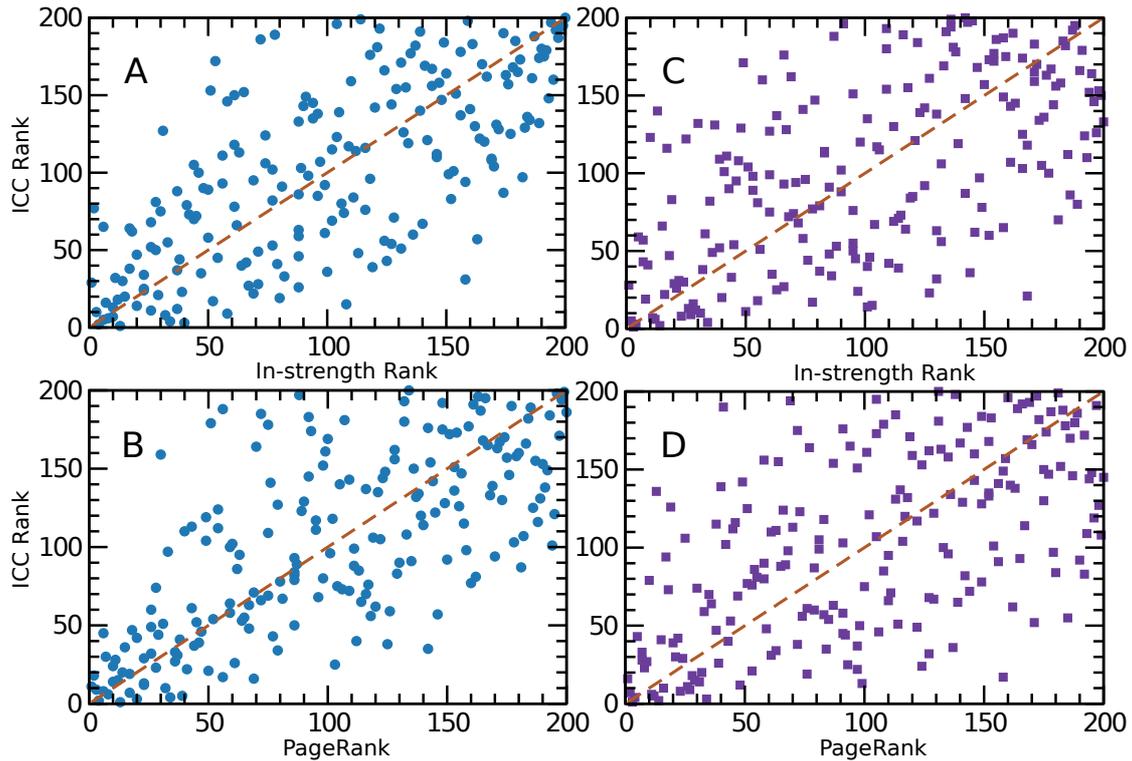}
\caption{ \label{fig:corr3}(Color online) (A) Scatter plot of between the rank positions obtained according to ICC points and In-strength rank for Test cricket ($1877-2011$) ; Spearman correlation $\rho$ = $0.69$.  (B) Scatter plot of between the rank positions obtained according to ICC points and PageRank score for Test cricket ($1877-2011$) ; Spearman correlation $\rho$ = $0.71$. (C) Scatter plot of between the rank positions obtained according to ICC points and In-strength rank for ODI cricket ($1971-2011$) ; Spearman correlation $\rho$ = $0.58$. (D) Scatter plot of between the rank positions obtained according to ICC points and PageRank score in ODI cricket ($1971-2011$) ; Spearman correlation $\rho$ = $0.59$.}
\end{center}
\end{figure*}

We find that our scheme provides sensible results that are in agreement with the points provided by ICC.{\bf The rankings provided by ICC take in account several factors like wickets taken, quality of pitch and opposition, match result etc. However, due to its opaqueness, ICC's methodology is incomprehensible. Our approach is both novel and transparent. For comparison, we choose the top $200$ bowlers according to ICC rankings \footnote{Since the information of ICC points is not consistently stored we choose the information of top $200$ bowlers in ODI and Test.}  and compare them with in-strength rank and PageRank. Figure~\ref{fig:corr3} shows that strong correlation exists between ranks obtained by network based tools and that provided by ICC}. This demonstrates that our network based approach captures the consensus opinions. 

Finally we propose a linear regression model for in-strength that takes into consideration known ranking schemes like PageRank, batting (bowling) average and ICC ranking,
\begin{equation}
s(i) = A_0 + A_1~\rho(i) + A_2~B_{Avg}(i) + A_3~\delta(i)  \,,
\label{outcome}
\end{equation}
where $s(i)$ is the in-strength, $\rho(i)$ is the PageRank of a player $i$. $B_{Avg}$ represents the batting (bowling) average of player $i$ and $\delta(i)$ is a dummy variable which takes the value $1$ if a player is placed in the top $100$ of ICC player ranking \cite{cricinfo}, and $0$ otherwise. As shown in Table~\ref{table_regression_1}, we observe that for bowlers in Test cricket ($1877-2011$), bowling average has no significant effect for in-strength, thus justifying the absence of correlation observed earlier in Figure~\ref{fig:corr2}(C, D).

\section{Conclusion}
To summarize, we quantified the performance of batsmen and bowlers in the history of cricket by studying the network structure of cricket players. Under the usual qualification of $2000$ balls bowled, {\it George Lohmann} emerge as the best bowler. Again, if we apply the qualification of at least $10$ dismissals, then {\it C. S. Marriott} is the best bowler. These constraints are arbitrary and hence gauging bowler's potential according to bowling average is not robust. The advantage of network analysis is that it doesn't introduce these `constraints' and yet provides consistent results. In such situation, in-strength and PageRank score stands out as an efficient measure of a bowler's ability. We would like to mention that although our study includes the 'quality' of bowling attack or 'quality' of dismissal of a batsman, we don't consider the fielding abilities or wicket-keeping abilities of the fielders. It is not possible to quantify the fielding ability of a fielder, other than by the number of catches, which is not a true measure of a fielder's ability. Some fielders are more athletic than others. Slip fielders always have a higher chance of taking a catch than others. Again, a batsman deserves more credit if he is able to beat athletic fielders like Jonty Rhodes, Ricky Ponting or Yuvraj Singh. Secondly, a bowler's ability is also judged by the nature of wicket. An excellent bowling performance on a batsman-friendly pitch holds greater merit than that on pitches which help bowlers. Similarly, scoring runs on difficult tracks always gets more attention than scoring runs on good batting tracks. In our analysis, due to non-availability of these informations, we didn't include these `external factors' in our analysis. 

Nevertheless a network based approach could  address the issue of relative performance of one player against other. Our study shows that SNA can indeed classify bowlers and batsmen based on the 'quality' of wickets taken or runs scored and not on the averages alone. Team selection is extremely important for any nation. SNA could be used as an objective way to aid the selection committee. A proper analysis of a player's domestic performance would help his(her) selection in the  national squad. Additionally, owners of the cash rich Indian Premier League (IPL) teams spend lots of money to hire players on a contract basis. The owners along with the coaches can identify talents based on the past `performance' of a player. Potentially our study could identify the greatest batsman of all time, based on a complete player-vs-player information, which at present we are unable to identify due to non-availability of data. Our analysis doesn't aim at replacing the existing system of ICC player ranking, which are based on expert opinions and has been optimized and almost perfected for many years. It serves as an alternate method to refine the existing ranking scheme of players and quantify the performance of a player. 

There are many additional features that could be included in the networks. For example, the networks in our analysis are static. A dynamic version of the network can be constructed following the ball-by-ball commentary and obtain a detailed analysis. Again, for batsmen there are players who score differently in different innings. There are leadership effects as well. Some players perform well under different skippers \footnote{The 1981 Ashes series where Ian Botham displayed tremendous performance under the inspiring leadership of Mike Brearley}. Bowlers are categorized into different categories based on their bowling style - pacers, medium pacers and spinners. Quantifying the `style' of bowling and effect of pitch conditions thus remains an open area of research. A rigorous analysis backed by a complete dataset of player-vs-player could very well answer the question - Was {\it Sir Don Bradman} the greatest ever ? In our quest to judge the most successful bowler in the history of cricket, one fact stands out : {\it M. Muralitharan} remains {\it il capo dei capi}.

\clearpage

\begin{table*}{\tiny                                                
\caption{\label{tab:table1} Ranking of top $50$ batsmen in Test cricket ($2001-2011$). We compare the rank of the batsmen according to their In-strength and compare them with the corresponding PageRank score, Batting average and best ever points according to ICC ratings.}
\begin{center}
\begin{tabular}{|c|c|c|c|c|c|c|}
\hline
\textbf{Rank} & \textbf{Batsman} & \textbf{Country} & \textbf{In strength} & \textbf{PageRank Score} & \textbf{Batting Average} & \textbf{ICC Points} \\
\hline
1 & K. C. Sangakkara & Sri Lanka & 131.520 & 0.189813 & 59.43 & 938 \\
2 & S. R. Tendulkar & India & 115.460 & 0.065442 & 55.13 & 898 \\
3 & R. T. Ponting & Australia & 113.582 & 0.049806 & 59.93 & 942 \\
4 & J. H. Kallis & South Africa & 103.545 & 0.030825 & 66.66 & 935 \\
5 & R. Dravid & India & 100.344 & 0.023313 & 54.31 & 892 \\
6 & V. Sehwag & India & 100.076 & 0.022095 & 51.87 & 866 \\
7 & D. P. M. D. Jayawardene & Sri Lanka & 99.131 & 0.022345 & 55.41 & 883 \\
8 & V. V. S. Laxman & India & 97.555 & 0.020722 & 49.58 & 781 \\
9 & S. Chanderpaul & West Indies & 96.319 & 0.019905 & 56.40 & 901 \\
10 & G. C. Smith & South Africa  & 88.943 & 0.014527 & 50.28 & 843 \\
11 & M. L. Hayden & Australia & 85.628 & 0.012232 & 56.27 & 935 \\
12 & Younis Khan & Pakistan & 83.255 & 0.011589 & 57.15 & 880 \\
13 & B. C. Lara & West Indies & 82.112 & 0.009571 & 60.88 & 991 \\
14 & A. N. Cook & England & 80.407 & 0.008708 & 48.69 & 836 \\
15 & A. J. Strauss & England &78.447 & 0.008470 & 41.60 & 769 \\
16 & K. P. Pietersen & England &77.312 & 0.007912 & 50.79 & 909 \\
17 & C. H. Gayle & West Indies &74.070 & 0.007991 & 43.27 & 755 \\
18 & A. B. de Villiers & South Africa  & 73.922 & 0.007441 & 51.00 & 776 \\
19 & M. E. K. Hussey & Australia &70.899 & 0.006557 & 51.29 & 921 \\
20 & M. P. Vaughan & England &65.216 & 0.005795 & 44.28 & 876 \\
21 & T. T. Samaraweera & Sri Lanka & 64.221 & 0.006355 & 60.08 & 750 \\
22 & J. L. Langer & Australia &62.221 & 0.005165 & 50.69 & 780 \\
23 & R. R. Sarwan & West Indies &58.056 & 0.005216 & 41.94 & 767 \\
24 & B. B. McCullum &  New Zealand &57.958 & 0.004427 & 36.90 & 673 \\
25 & D. L. Vettori &  New Zealand & 57.919 & 0.006190 & 35.70 & 672 \\
26 & M. J. Clarke & Australia &57.830 & 0.004501 & 50.43 & 855 \\
27 & H. H. Gibbs & South Africa  & 57.566 & 0.004485 & 46.67 & 825 \\
28 & I. R. Bell &England & 57.356 & 0.004460 & 47.80 & 822 \\
29 & M. E. Trescothick & England &55.499 & 0.003753 & 45.83 & 818 \\
30 & T. M. Dilshan & Sri Lanka & 54.190 & 0.005119 & 44.37 & 700 \\
31 & A. C. Gilchrist & Australia &53.751 & 0.003999 & 48.16 & 874 \\
32 & D. R. Martyn & Australia &53.141 & 0.003723 & 48.32 & 848 \\
33 & H. M. Amla & South Africa  & 52.579 & 0.003863 & 48.52 & 842 \\
34 & A. Flintoff & England &48.845 & 0.003620 & 34.06 & 645 \\
35 & Inzamam ul Haq & Pakistan & 46.838 & 0.003317 & 57.75 & 870 \\
36 & S. T. Jayasuriya & Sri Lanka & 46.401 & 0.003044 & 42.66 & 770 \\
37 & S. M. Katich & Australia &45.676 & 0.003037 & 46.01 & 807 \\
38 & S. C. Ganguly & India & 45.418 & 0.002945 & 42.26 & 713 \\
39 & M. V. Boucher & South Africa  & 44.699 & 0.004131 & 31.78 & 566 \\
40 & L. R. P. L. Taylor & New Zealand & 44.060 & 0.002538 & 45.72 & 775 \\
41 & G. Gambhir & India &43.806 & 0.002713 & 47.51 & 886 \\
42 & P. D. Collingwood & England &43.739 & 0.002763 & 40.57 & 730 \\
43 & S. P. Fleming &  New Zealand & 43.374 & 0.002823 & 44.15 & 725 \\ 
44 & M. S. Dhoni & India & 43.344 & 0.002406 & 37.84 & 662 \\
45 & M.  S. Atapattu & Sri Lanka & 40.912 & 0.002540 & 44.72 & 670 \\
46 & A. G. Prince & South Africa  & 40.704 & 0.002530 & 43.12 & 756 \\
47 & Habibul Bashar & Bangladesh & 40.702 & 0.002456 & 31.03 & 656 \\
48 & Mohammad Ashraful & Bangladesh & 38.937 & 0.002942 & 22.62 & 491 \\
49 & M. J. Prior & England &  38.594 & 0.001972 & 46.75 & 679 \\
50 & Imran Farhat & Pakistan & 37.910 & 0.002399 & 33.03 & 575 \\
\hline
\end{tabular}
\end{center}
}
\end{table*}

\clearpage

\begin{table*}{\tiny                                               
\caption{\label{tab:table2} Ranking of top $50$ batsmen in ODI cricket ($1999-2011$). We compare the rank of the batsmen according to their In-strength and compare them with the corresponding PageRank score, Batting average and best ever points according to ICC ratings.}
\begin{center}
\begin{tabular}{|c|c|c|c|c|c|c|}
\hline
\textbf{Rank} & \textbf{Batsman} & \textbf{Country} & \textbf{In strength} & \textbf{PageRank Score} & \textbf{Batting Average} & \textbf{ICC Points} \\
\hline
1 & K. C. Sangakkara & Sri Lanka & 128.075 & 0.165704 & 42.59 & 863 \\
2 & R. T. Ponting & Australia & 127.058 & 0.095677 & 46.94 & 829 \\
3 & S. R. Tendulkar &  India & 120.251 & 0.052469 & 50.90 & 898 \\
4 & D. P. M. D. Jayawardene & Sri Lanka & 115.475 & 0.040357 & 38.33 & 738 \\
5 & Yuvraj Singh &  India & 109.620 & 0.027228 & 40.48 & 787 \\
6 & V. Sehwag &  India & 104.183 & 0.022008 & 38.51 & 774 \\
7 & J. H. Kallis & South Africa  &97.150 & 0.016652 & 49.89 & 817 \\
8 & M. S. Dhoni &  India & 96.639 & 0.014579 & 56.44 & 836 \\
9 & Younis Khan & Pakistan &  90.578 & 0.013467 & 37.19 & 659 \\
10 & S. T. Jayasuriya & Sri Lanka & 89.352 & 0.012719 & 36.13 & 838 \\
11 & G. C. Smith & South Africa  &88.873 & 0.011473 & 40.25 & 784 \\
12 & M. J. Clarke & Australia & 86.790 & 0.010249 & 51.50 & 750 \\
13 & R. Dravid &  India & 85.407 & 0.009736 & 48.95 & 749 \\
14 & A. C. Gilchrist & Australia & 79.554 & 0.007398 & 36.95 & 820 \\
15 & C. H. Gayle & West Indies & 78.427 & 0.008268 & 42.50 & 804 \\
16 & S. Chanderpaul &  West Indies & 77.500 & 0.008227 & 48.10 & 776 \\
17 & M. E. K. Hussey & Australia & 77.276 & 0.006517 & 53.15 & 857 \\
18 & M. L. Hayden &Australia &  76.883 & 0.007100 & 46.95 & 850 \\
19 & T. M. Dilshan & Sri Lanka & 71.624 & 0.006672 & 38.74 & 765 \\
20 & H. H. Gibbs & South Africa  &71.192 & 0.006471 & 40.06  & 750 \\
21 & B. B. McCullum & New Zealand & 67.281 & 0.005514 & 31.98 & 664 \\
22 & S. C. Ganguly &  India & 65.523 & 0.004433 & 41.73 & 844 \\
23 & P. D. Collingwood & England &64.758 & 0.004741 & 39.50 & 697 \\
24 & S. B. Styris & New Zealand & 64.627 & 0.005155 & 37.69 & 663 \\
25 & Shoaib Malik & Pakistan &  64.263 & 0.005298 & 38.90 & 685 \\
26 & R. R. Sarwan & West Indies &  63.823 & 0.004680 & 48.99 & 780 \\
27 & G. Gambhir &  India & 62.207 & 0.004319 & 44.93 & 722 \\
28 & A. B. de Villiers & South Africa & 61.920 & 0.004008 & 55.53 & 803 \\
29 & W. U. Tharanga & Sri Lanka & 60.773 & 0.003835 & 37.38 & 663 \\
30 & A. J. Strauss & England &60.650 & 0.003881 & 37.29 & 698 \\
31 & M. S. Atapattu & Sri Lanka & 60.328 & 0.003837 & 44.77 & 738 \\
32 & Shahid Afridi & Pakistan &  58.139 & 0.004976 & 24.39 & 663 \\
33 & S. P. Fleming & New Zealand & 54.770 & 0.003231 & 36.20 & 697 \\
34 & Inzamam ul Haq & Pakistan &  53.965 & 0.003072 & 40.68 & 801 \\
35 & K. P. Pietersen & England & 53.804 & 0.003069 & 43.54 & 833 \\
36 & Yousuf Youhana & Pakistan &  53.255 & 0.003473 & 52.36 & Not Available \\
37 & M. E. Trescothick & England & 52.613 & 0.002360 & 40.48 & 797 \\
38 & M. V. Boucher & South Africa  &52.510 & 0.003781 & 32.72 & 621 \\
39 & S. K. Raina &  India & 51.335 & 0.002705 & 37.90 & 658 \\
40 & Abdul Razzaq & Pakistan &  49.876 & 0.003503 & 35.05 & 328 \\
41 & S. R. Watson & Australia & 49.387 & 0.003247 & 43.70 & 773 \\
42 & A. Symonds & Australia & 49.098 & 0.002789 & 46.49 & 776 \\
43 & I. R. Bell & England & 46.834 & 0.002560 & 38.96 & 702 \\
44 & C. D. McMillan & New Zealand & 43.952 & 0.002434 & 30.86 & 648 \\
45 & V. Kohli & India & 42.868 & 0.001593 & 55.23 & 799 \\
46 & Salman Butt & Pakistan &  42.253 & 0.001602 & 44.66 & 683 \\
47 & Shakib Al Hasan & Bangladesh & 41.856 & 0.002256 & 36.31 & 659 \\
48 & H. M. Amla & South  Africa & 40.457 & 0.001523 & 65.48 & 886 \\
49 & B. R. M. Taylor & New Zealand & 40.289 & 0.002036 & 36.57 & 654 \\
50 & Tamim Iqbal & Bangladesh & 39.593 & 0.001845 & 33.77 & 629 \\
\hline
\end{tabular}
\end{center}
}
\end{table*}

\clearpage

\begin{table}{\tiny                                                
\caption{\label{tab:table3} Ranking of top $50$ bowlers in the history of Test cricket ($1877-2011$). We compare the rank of the bowlers according to their In-strength and compare them with the corresponding PageRank score, Batting average and best ever points according to ICC ratings.}
\begin{center}
\begin{tabular}{|c|c|c|c|c|c|c|}
\hline
\textbf{Rank} & \textbf{Bolwers} & \textbf{Country} & \textbf{In strength} & \textbf{PageRank Score} & \textbf{Bowling Average} & \textbf{ICC Points} \\
\hline
1  &  M. Muralitharan & Sri Lanka & 1838.727 & 0.081376 & 22.72& 920\\
2  &  S. K. Warne & Australia & 1600.098 & 0.037871 &  25.41& 905\\
3  &  G. D. McGrath& Australia & 1581.467 & 0.035376 & 21.64& 914\\
4  &  A. Kumble & India & 1207.115 & 0.028108 & 29.65& 859\\
5  &  C. A. Walsh & West Indies &  1206.669 & 0.028407 & 24.44 &867 \\
6  &  C. E. L. Ambrose& West Indies & 1118.653 & 0.014483 & 20.99 & 912\\
7  &  M. D. Marshall& West Indies& 1077.349 & 0.027349 & 20.94 & 910\\
8   & S. M. Pollock& South Africa & 1060.700 & 0.008220 &  23.11& 909\\
9 & R. J. Hadlee & New Zealand & 1045.247 & 0.020100 & 22.29 & 909 \\
10   & D. K. Lillee& Australia & 907.015 & 0.011724 &  23.92& 884\\
11  &  Wasim Akram& Pakistan & 906.455 & 0.007559 &  23.62& 830\\
12   & Imran Khan& Pakistan &  891.679 & 0.012749 & 22.81& 922\\
13   & A.A. Donald& South Africa & 842.499 & 0.003900 &  22.25& 895\\
14   & M. Ntini& South Africa & 836.285 & 0.004674 &  28.82& 863\\
15  &  Waqar Younis& Pakistan & 832.806 & 0.004918 &  23.56 & 909\\
16   & F. S. Trueman& England & 791.479 & 0.034600 &  21.57& 898\\
17  &  N Kapil Dev& India & 778.960 & 0.006425 & 29.64 & 877\\
18   & Harbhajan Singh& India & 761.382 & 0.004886 &  32.22&765 \\
19   & I. T. Botham& England & 720.371 & 0.004315 &  28.40& 911\\
20  &  R. G. D.  Willis&England & 719.321 & 0.005895 &  25.20& 837\\
21  &  D. L. Underwood& England& 697.950 & 0.008028 &  25.83&907 \\
22  &  W. P. U. J. C. Vaas& Sri Lanka & 668.274 & 0.003754 &  29.58& 800\\
23   & D. W. Steyn& South Africa & 663.894 & 0.003566 &  23.07& 902\\
24   & J. Garner& West Indies & 647.740 & 0.002803 &  20.97& 890\\
25   & B. Lee& Australia & 624.158 & 0.002397 & 30.81 & 811\\
26   & M. A. Holding& West Indies & 615.905 & 0.003025 &  23.68& 860\\
27   & L. R. Gibbs& West Indies &  607.816 & 0.010326 & 29.09& 897\\
28   & R. R. Lindwall& Australia & 593.348 & 0.008941 &  23.03& 897\\
29   & C. J. McDermott& Australia & 590.881 & 0.002318 &  28.63& 794\\
30   & J. N. Gillespie& Australia & 585.951 & 0.002121 & 26.13 & 812\\
31   & J. B. Statham& England & 575.871 & 0.007935 &  24.84& 810\\
32   & S. F. Barnes& England & 575.551 & 0.011649 &  16.43& 932\\
33   & Z Khan& India & 574.541 & 0.003255 &  31.78& 752\\
34   & A. V. Bedser&England & 573.140 & 0.006187 & 24.89 &903 \\
35   & D. L. Vettori& New Zealand & 558.336 & 0.003616 &  33.65& 681\\
36   & A. K. Davidson& Australia & 531.038 & 0.004510 & 20.53 & 908\\
37  &  M. J. Hoggard& England & 523.946 & 0.001646 &  30.56& 795\\
38   & J. C. Laker& England & 522.186 & 0.004353 &  21.24& 897\\
39  &  G. D. McKenzie& Australia & 518.735 & 0.003349 &  29.78& 846\\
40   & Saqlain Mushtaq& Pakistan & 513.114 & 0.001625 &  29.83& 771\\
41  &  R. Benaud& Australia & 512.006 & 0.003863 &  27.03& 863\\
42   & C. V. Grimmett& Australia & 509.586 & 0.024239 & 24.21 & 901\\
43   & J. H. Kallis& South Africa & 500.176 & 0.003184 &  32.51&742 \\
44   & Mohammad Asif & Pakistan & 499.581 & 0.001268 &  24.36& 818 \\
45   & B. S. Bedi& India & 488.933 & 0.002868 &  28.71& 804\\
46   & J. M. Anderson& England & 486.732 & 0.002245 & 30.46 &813 \\
47   & A. R. Caddick& England& 483.068 & 0.001447 & 29.91 & 732\\
48   & K. R. Miller& Australia & 476.808 & 0.003903 & 22.97 & 862 \\
49   & J. A. Snow& Australia & 468.001 & 0.002138 &   26.66 & 835 \\
50   & D. Gough& England & 457.295 & 0.001287 & 28.39& 794\\

\hline
\end{tabular}
\end{center}
}

\end{table}

\clearpage

\begin{table}{\tiny                                            
\caption{\label{tab:table4} Ranking of top $50$ bowlers in the history of ODI cricket ($1971-2011$). We compare the rank of the bowlers according to their In-strength and compare them with the corresponding PageRank score, Batting average and best ever points according to ICC ratings.}
\begin{center}
\begin{tabular}{|c|c|c|c|c|c|c|}
\hline
\textbf{Rank} & \textbf{Bowlers} & \textbf{Country} & \textbf{In strength} & \textbf{PageRank Score} & \textbf{Bowling Average} & \textbf{ICC Points} \\
\hline
1  &  M. Muralitharan& Sri Lanka & 607.375 & 0.170207 & 23.08 & 913\\
2  &  Wasim Akram& Pakistan & 601.274 & 0.111784 & 23.52 &850 \\
3   & G. D. McGrath& Australia &  473.596 & 0.029389 & 22.02&903 \\
4   & Waqar Younis& Pakistan &  471.019 & 0.030567 & 23.84& 778\\
5   & S. M. Pollock& South Africa&  440.701 & 0.018813 & 24.50& 917\\
6   & B. Lee& Australia & 437.882 & 0.020709 & 23.18 & 852\\
7   & W. P. U. J. C. Vaas & Sri Lanka & 426.005 & 0.019129 &  27.53& 860\\
8   & Saqlain Mushtaq& Pakistan& 381.207 & 0.011874 &  21.78& 804\\
9   & A. A. Donald& South Africa & 331.312 & 0.011041 &  21.78& 794\\
10  &  M. Ntini& South Africa & 305.877 & 0.007624 &  24.65& 783\\
11   & J. Srinath& India &  305.067 & 0.008372 & 28.08& 742\\
12   & S. K. Warne& Australia & 296.573 & 0.007119 &  25.73& 786\\
13   & A. Kumble& India &  293.592 & 0.009605 & 30.89& 797\\
14   & A. B. Agarkar&India & 283.160 & 0.005718 &  27.85& 675\\
15  &  Shahid Afridi& Pakistan &  281.853 & 0.007799 & 33.37&623 \\
16   & D. L. Vettori& New Zealnd& 266.683 & 0.006241 &  31.48& 788\\
17   & Z. Khan& India &  262.253 & 0.006112 & 29.03& 700\\
18  &  Harbhajan Singh& India & 261.937 & 0.005727 &  33.40& 735\\
19   & C. E. L. Ambrose& West Indies &  259.694 & 0.005700 & 24.12& 877\\
20   & D. Gough& England & 247.125 & 0.004335 & 26.42 & 767\\
21   & S. T. Jayasuriya& Sri Lanka & 236.641 & 0.006771 &  36.75& 591\\
22  &  N Kapil Dev&India & 234.467 & 0.009698 &  27.45& 845\\
23   & J. H. Kallis&South Africa & 234.421 & 0.005161 & 31.69 &641 \\
24   & Abdul Razzaq& Pakistan &  234.380 & 0.004718 & 31.83& 678\\
25   & K. D. Mills& New Zealand &  218.920 & 0.003573 & 25.94& 722\\
26  &  C. J. McDermott& Australia & 212.171 & 0.003862 & 24.71&808 \\
27  &  H. H. Streak& Zimbabwe &  211.982 & 0.003212 & 29.82& 717\\
28   & J. Garner& West Indies & 209.613 & 0.006778 & 18.84 &940 \\
29   & S. E. Bond& New Zealand & 208.962 & 0.002790 & 20.88 & 809\\
30   & C. A. Walsh& West Indies & 203.122 & 0.004218  &  30.47& 801\\
31   & N. W. Bracken& Australia &  202.785 & 0.002428 & 24.36& 805\\
32  &  C. L. Cairns& New Zealand & 197.498 & 0.003479 & 32.80& 784\\
33   & A. Flintoff& England & 192.269 & 0.002707 & 24.38& 755\\
34   & J. M. Anderson& England & 191.432 & 0.003190 & 30.83 & 687\\
35  &  M. G. Johnson& Australia & 187.358 & 0.002140 &  25.22& 724\\
36  &  C. R. D. Fernando&Sri Lanka & 186.151 & 0.003019 &  30.20&624 \\
37 & R. J. Hadlee & New Zealand & 182.408 & 0.005386 & 21.56 & 923\\
38   & B. K. V. Prasad& India&  177.538 & 0.000349 & 32.30& 692\\
39   & Imran Khan& Pakistan & 174.633 & 0.005972 &  26.61& 780\\
40  &  L. Klusener& South Africa & 174.271 & 0.000358 & 29.95& 657\\
41   & Abdur Razzak& Bangladesh & 173.970 & 0.002919 &  28.12& 675 \\
42   & M. A. Holding& West Indies & 160.605 & 0.004294 &  21.36& 875\\
43   & C. Z. Harris& New Zealand & 159.101 & 0.002170 &  37.50& 659\\
44  &  M. D. Marshall& West Indies & 158.326 & 0.003466 &  26.96&891 \\
45   & S. C. J. Broad& England & 158.194 & 0.000867 &  26.95&701 \\
46   & C. L. Hooper& West Indies& 154.591 & 0.002299 & 36.05&679 \\
47  &  S. L. Malinga& Sri Lanka & 154.017 & 0.002022 & 26.35 & 674\\
48   & J. N. Gillespie& Australia & 150.864 & 0.002121 &  25.42& 823\\
49   & G. B. Hogg& Australia & 149.910 & 0.000216 &  26.84& 688\\
50   & I. K. Pathan& India & 148.536 & 0.000352 &  29.89& 722\\
\hline
\end{tabular}
\end{center}
}
\end{table}

\clearpage

\begin{table}\small
\centering
\caption{{\bf Results for the linear regression} We mark in bold font the coefficients that are statistically significant (p-value$<0.05$). }
\begin{tabular}{lc|ccc}

 &&&\\

         Number of Bowlers  & Test & $1877-2011$& ~2616 \\
	Number of Bowlers  & ODI & $1971-2011$& ~1914 \\
	Number of Batsmen  & Test & $2001-2011$& ~599 \\ 
	Number of Batsmen & ODI & $1999-2011$& ~1027 \\
&&&\\
&&&\\
&&& \bf{Model}\\
&&&&\\
&  &Coef.&  Std.Err.  & p-value

\\ \hline
	
        \bf{Model for $s$} for Bowlers in Test ($1877-2011$) & &\\
       
&&&\\

	~~~Intercept & &\bf{-8.16} &1.12   &$<1\times 10^{-16}$       \\ 
	~~~Pagerank&$\rho$ & \bf{ 128093.4} &1324.76 &$<1\times 10^{-16}$   \\ 	
	~~~Bowling Average&$B_{Avg}$ & 0.042 & 0.025 &$0.091$   \\ 
	~~~Dummy&$\delta$ & \bf{ 24.05} & 1.12 &$<1\times 10^{-7}$   \\ 

&&&\\

        ~~~R-squared& &$0.8843$\\ 

&&&\\      
 \hline

         \bf{Model for $s$} for Bowlers in ODI($1971-2011$) & &\\
       
&&&\\

	~~~Intercept & &\bf{ -2.41} &0.438   &$<1\times 10^{-16}$       \\ 
	~~~Pagerank&$\rho$ & \bf{37766.47} &  312.61 &$<1\times 10^{-16}$   \\ 	
	~~~Bowling Average&$B_{Avg}$ & \bf{0.036} & 0.011 &$0.001$   \\ 
	~~~Dummy&$\delta$ & \bf{16.98} & 1.73 &$<1\times 10^{-16}$   \\ 

&&&\\

        ~~~R-squared& &$0.9265$\\ 

&&&\\      
 \hline

         \bf{Model for $s$} for Batsmen in Test ($2001-2011$) & &\\
       
&&&\\

	~~~Intercept & &\bf{-1.78} &0.822   &$0.031$       \\ 
	~~~Pagerank&$\rho$ & \bf{825.39} & 50.37 &$<1\times 10^{-16}$   \\ 	
	~~~Batting Average&$B_{Avg}$ & \bf{ 0.289} & 0.036 &$<1\times 10^{-16}$   \\ 
	~~~Dummy&$\delta$ & \bf{26.55} & 1.496 &$<1\times 10^{-16}$   \\ 

&&&\\

        ~~~R-squared& &$0.7201$\\ 

&&&\\      
 \hline 

         \bf{Model for $s$} for Batsmen in ODI ($1999-2011$) & &\\
       
&&&\\

	~~~Intercept & & -.029 &0.53   &$0.956$       \\ 
	~~~Pagerank&$\rho$ & \bf{1005.56} & 48.39 &$<1\times 10^{-16}$   \\ 	
	~~~Batting Average&$B_{Avg}$ & \bf{ 0.159} & 0.022 &$<1\times 10^{-16}$   \\ 
	~~~Dummy&$\delta$ & \bf{30.122} & 1.203 &$<1\times 10^{-16}$   \\ 

&&&\\

        ~~~R-squared& &$0.6627$\\ 

&&&\\      	
\end{tabular}
\label{table_regression_1}
\end{table}


\clearpage
 
\begin{thebibliography}{}

\bibitem{laszlo} R. Albert and A. Barabasi, \emph{ Reviews of Modern Physics} {\bf 74} (2002), B. Tadic, G.J. Rodgers, S. Thurner, \emph{ Int. J. Bifurcation and Chaos } {\bf 17}, 7, 2363 (2007)

\bibitem{freeman} Freeman, L.~C., Borgatti,  S.~P., White, D.~R., \emph{Social Networks} {\bf 13}(2), 141-154 (1991)

\bibitem{moviewatts} D.~J.~Watts and S.~H.~Strogatz \emph{Nature}, 393, 440. (1998)

\bibitem{castellano} Castellano C, Fortunato S, Loreto V \emph{Reviews of Modern Physics}, 81, 591 (2009)

\bibitem{newman8} M.~E.~J.~Newman \emph{Proc. Nat. Acad. Sci. USA}, 98, 404. (2001)

\bibitem{newman9} M.~E.~J.~Newman \emph{Phys. Rev. E}, 64, 016132. (2001)

\bibitem{pan1} R.~K.~Pan and J.~Saramaki \emph{Europhysics Letters}, 97, 18007 (2012)

\bibitem{pan2} R.~K.~Pan, K.~Kaski and S.~Fortunato \emph{Scientific Reports} 2, 902 (2012)

\bibitem{solla} de Solla Price DJ,  \emph{Science}, 149, 510 (1965)

\bibitem{redner} Chen P, Xie H, Maslov S, Redner S \emph{Journal of Informetrics}, 1, 8 (2007)

\bibitem{bergstrom} Bergstrom CT, West J \emph{Neurology}, 71, 1850 (2008)

\bibitem{bergstrom2} West J, Bergstrom T, Bergstrom CT \emph{Journal of American Society of Information Science and Technology}, 61, 1800 (2010)

\bibitem{vespignani}Radicchi F, Fortunato S, Markines B, Vespignani A \emph{Phys Rev E}, 80, 056103 (2009)

\bibitem{naim05} Ben-Naim, E., F.~Vazquez, and S.~Redner \emph{J. Korean Phys. Soc.}, 50, 124. (2007)

\bibitem{bittner} Bittner E, Nussbaumer A, Janke W, Weigel M \emph{European Physical Journal B} 67, 459 (2009) 

\bibitem{petersen08} Petersen, A.~M., W.~S. Jung, and H.~E. Stanley \emph{EPL}, 83, 50010. (2008)

\bibitem{sire09} Sire, C. and S.~Redner, 67, 473--481\emph{Eur. Phys. J. B}  (2009) 

\bibitem{naim07} Ben-Naim, E., S.~Redner, and F.~Vazquez \emph{EPL}, 77, 30005. (2007)

\bibitem{skinner10} Skinner, B. \emph{Journal of Quantitative Analysis in Sports}, 6, 3. (2010)

\bibitem{guerra}Guerra YD, Gonzalez JMM, Montesdeoca SS, Ruiz DR, Garcia-Rodriguez A, GarciaManso JM \emph{Physica A} 391, 2997 (2012)

\bibitem{Rubner} Heuer A, Rubner O \emph{European Physical Journal B } 67, 445 (2009) 

\bibitem{Ribeiro} Ribeiro HV, Mendes RS, Malacarne LC, Picoli S, Santoro PA \emph{European Physical Journal B} 75, 327 (2010) 

\bibitem{bhandari} Bhandari, I., Colet, E., Parker, J., Pines, Z., Pratap, R.,  Ramanujam, K.  \emph{Mining and Knowledge Discovery} 1, 121 (1997)

\bibitem{condon} Condon, E. M., Golden, B. L., Wasil, E. A.  \emph{Computers and Operations Research} 26(13), 1243 (1999)

\bibitem{sharda} S.~R.~Iyer, and R.~Sharda \emph{Expert Systems with Applications}, 36, 5510 (2009)

\bibitem{wilson1995} Wilson, R.  \emph{Interfaces} 25(4), 44 (1995)

\bibitem{majer} Maier, K. D., Wank, V.,  Bartonietz, K.  \emph{Sports Engineering} 3(1), 57 (2000)

\bibitem{jurgen}J{\"u}rgen, P., Arnold, B. \emph{Proceedings of the 8th annual congress of the European college of sport science} (pp. 342) Salzburg: ECSS (2003) 


\bibitem{heuer10} Heuer, A., C.~M\"uller, and O.~Rubner \emph{EPL}, 89, 38007. (2010)

\bibitem{malacarne} R.~S.~Mendes, L.~C.~Malacarne and C.~Anteneodo \emph{European Physical Journal B}, 57, 357. (2007)

\bibitem{duch10} Duch, J., J.~S. Waitzman, and L.~A.~N. Amaral \emph{PLoS ONE}, 5, e10937. (2010)

\bibitem{yokoyama} Yamamoto, Y., and Yokoyama, K. \emph{PLoS ONE}, 6(12), (2011)

\bibitem{onody04} Onody, R.~N. and P.~A. de~Castro \emph{Phys. Rev. E}, 70, 037103. (2004)

\bibitem{mendes2011} P.~Passos, K.~Davids, D.~Araujo, N.~Paz, J.~Minguens and J.~Mendes \emph{Journal of Science and Medicine in Sport}, 14, 170 (2011) 

\bibitem{saavedra09} Saavedra, S., S.~Powers, T.~McCotter, M.~A. Porter, and P.~J. Mucha \emph{Physica A} 389, 1131 (2009)

\bibitem{newman2005} Park, J. and M.~E.~J. Newman \emph{Journal of Statistical Mechanics : Theory and Experiment}, 10. (2005)

\bibitem{radicchi11} Radicchi, F. \emph{PloS ONE}, 6, e17249. (2011)

\bibitem{mukherjee2012} S. Mukherjee, \emph{Physica A} 391, 6066 (2012) 

\bibitem{Lusher} D.~Lusher, G.~Robins \& P.~Kremer \emph{ Measurement in Physical Education and Exercise Science}, 14, 211 (2010)

\bibitem{amy2007} Amy, M. \emph{The history of cricket}, eSSORTMENT, $http://www.essortment.com/hobbies/historycricket_sngj.htm.$ (2007)

\bibitem{Vani2010} Borooah, V.~K. and J.~E. Mangan  \emph{Journal of Quantitative Analysis in Sports}, 6. (2010)

\bibitem{lemmer} Lemmer, H.~.H. \emph{Journal of Sports Sciences and Medicine}, 10, 630 (2011)

\bibitem{cricinfo} www.espncricinfo.com

\bibitem{danila} B. Danila, Y. Yu, S. Earl, J. A. Marsh, Z. Toraczkai and K. E. Bassler, \emph{Phys. Rev. E}, {\bf 74}, 046114 (2006)


\bibitem{jamming} Z. Toroczkai, K. E.Bassler, \emph{Nature}, {\bf 428} (2004)

\bibitem{brinpage} Brin, S. and Page, L \emph{Computer Networks ISDN Syst.} 30, 107, (1998)

\bibitem{easley} David Easley and Jon Kleinberg, \emph{Networks, Crowds, and Markets : Reasoning about a Highly Connected World}, \bf{Cambridge University Press} (2010)

\end{thebibliography}
\end{document}